\documentclass[twocolumn]{aastex631}
\usepackage{graphicx} 
\usepackage{amsmath}
\usepackage[]{natbib}
\usepackage{color}

\graphicspath{{./}}

\begin{document}

\title{Morphological evolution of disk galaxies and their concentration, asymmetry and clumpiness (CAS) properties in simulations across Toomre's $Q$ parameter}
\author{T. Chantavat}
\affiliation{The Institute for Fundamental Study, Naresuan University, Phitsanulok, 65000, Thailand}
\author[0000-0002-0269-0135]{S. Yuma}
\author{P. Malelohit}
\affiliation{Department of Physics, Faculty of Science, Mahidol University, Bangkok 10400, Thailand}
\author{T. Worrakitpoonpon}
\affiliation{Institute of Science, Suranaree University of Technology, Nakhon Ratchasima 30000, Thailand}

\correspondingauthor{T. Worrakitpoonpon}
\email{worraki@gmail.com}
\date{Received <date> / accepted <date>}
\date{}


\begin{abstract}
    We investigate the morphological and structural evolutions of disk galaxies in simulations for a wide range of Toomre's $Q$ parameter. In addition to the inspection of conventional bar modes, we compute the concentration, asymmetry and clumpiness (CAS) parameters to enlarge the understanding of the galaxy evolution. These parameters are widely employed to analyze the light distribution of the observed galaxies, but the adaptation to numerical simulations is not much considered. While the bar formation takes place in a considerable range of $Q$ around $1$, barred galaxies originating from $Q>1$ and $Q<1$ disks yield the CAS values that differ significantly. Disks starting with $Q<1$ develop clumps due to local gravitational instabilities along with the bar and these clumps play a central role in enhancing the CAS values. That process is absent in $Q>1$ counterparts in which the evolution is dominated by linearly unstable two-armed modes that lead to lower CAS values. Likewise, unbarred galaxies that are obtainable from disks with $Q$ far below and far above $1$ exhibit greatly different CAS magnitudes. It turns out that the CAS parameters can serve as indicators of the initial kinematical state and the evolution history of a disk of any morphology. In addition, we find an alternative mechanism of the formation of the lopsided barred galaxy when $Q\lesssim 1$. Bars that evolve in the midst of the clumps can spontaneously become lopsided at the end.
\end{abstract}

\keywords{Galaxy dynamics(591) --- Barred spiral galaxies(136) --- Galaxy formation(595) --- N-body simulations(1083)}

\section{Introduction}

Disk galaxies constitute an important fraction in the Hubble sequence and their morphologies manifest great diversities such as spiral, barred-spiral, and lenticular shapes. However, a full understanding of their origins remains to be resolved. In the standard theoretical framework, sub-structures harbored in the disk components are considered as the amplified perturbations of various forms. For instance, the galactic bar is the remnant of the preceding linearly unstable two-armed modes \citep{kalnajs_1971,contopoulos+papayannopoulos_1980,vauterin+dejonghe_1996,jalali+hunter_2005}. On the other hand, the spiral structures can be achieved by the spiral density wave \citep{lin+shu_1964}, the sheared gravitationally unstable medium \citep{goldreich+lynden_bell_1965}, or the sheared synchronized epicyclic motion \citep{julian+toomre_1966,toomre_1981,de_rijcke_et_al_2019}. 

Despite the well-formulated theoretical frameworks, numerical simulations unveiled much intricacy as it was found that the final disk morphologies depended strongly on the disk and halo intrinsic properties. A number of simulations demonstrated that the Toomre's $Q$ parameter \citep{toomre_1964} could determine how a circular disk morphologically evolved \citep{hohl_1971,athanassoula+sellwood_1986,curir_et_al_2006,bekki_2023,worrakitpoonpon_2023}. Otherwise, the disk random kinetic energy was proved usable as an indicator of the fate of a disk  \citep{op_1973,eln_1982,athanassoula_2008let,romeo_et_al_2023}. In a larger scope, the concern of the disk stability was not only limited to the disk initial kinematical properties, but the halo physical and kinematical properties were also found important to the subsequent disk evolution. Many works reported the correlation between the halo/bulge mass concentration and the stability against bar modes \citep{sellwood_1980,sellwood+evans_2001,athanassoula+misiriotis_2002,shen+sellwood_2004,athanassoula_et_al_2005,sheth_et_al_2008,sellwood_2016cmc,saha+elmegreen_2018,kataria+das_2018,kataria+das_2019,kataria_et_al_2020,jang+kim_2023}. The halo spin was also found to be influential to the disk dynamical evolution \citep{saha+naab_2013,long_et_al_2014,collier_et_al_2019,kataria+shen_2022,li_et_al_2023,joshi+widrow_2024,chiba+kataria_2024}. On the other hand, the formation of the spiral structure rather relied on local effects such as the local instabilities and the disk shearing \citep{d_onghia_et_al_2013,fujii_et_al_2018,michikoshi+kokubo_2020}. The cold disk environment was the favorable condition to form spiral arms \citep{evans+read_1998,zakharova_et_al_2023}, whereas the disk thickness tended to suppress that process \citep{ghosh+jog_2018,bauer+widrow_2019}. Unlike the bar component, the formation and evolution of the spiral structures were less dependent on the choices of halo parameters \citep{sellwood_2021spiral}.

In the observational aspect, morphological studies of galaxies started from basic visual classification to the classification via physical properties of the galaxies. The classic Hubble sequence classified galaxies into elliptical, spiral, and irregular galaxies based on the apparent stellar distribution of galaxies \citep{hubble1926}. 
The visual inspection, however, has limitations for distant galaxies, as their surface brightness drops due to the cosmological dimming effect. Feature like spiral arms may get too faint to be detected at high redshifts \citep[e.g., ][]{papaderos2023}. 
In fact, distant star-forming galaxies do not share the same structure or morphology as the local spiral galaxies \citep{yuma2011, yuma2012}. Star-forming galaxies started to show a similar structure as those in the local universe at $z<0.85$ \citep{takeuchi2015}. 

Another approach is to fit the azimuthally averaged surface brightness profile with a predesignate formula such as de Vaucouleurs, exponential, and S\'ersic profiles \citep{devaucouleurs1948, sersic1963}. 
This method is efficiently used to distinguish 
an elliptical from the spiral galaxy in the local universe \citep[e.g.,][]{fischer2019,SDSS2022} up to very high redshifts of $z=16$ 
\citep{ono2023}. 
However, bulge-disc decomposition by fitting two components of S\'ersic and exponential profiles to the galaxies is sometimes physically unrealistic with the S\'ersic index of the bulge as high as $n\sim8$ \citep{fischer2019}. 

Non-parametric measurement of light distribution in a galaxy has been intensively introduced to diminish the effect of model or assumption \citep[e.g.,][]{bershady2000, conselice2000, conselice2003, conselice2014}. Concentration ($C$), 
asymmetry ($A$), and clumpiness ($S$) indices or so called CAS system are among 
the most common non-parametric methods used to study the galaxy structure. 
Besides their efficiency in classifying elliptical and spiral galaxies, 
the $CAS$ parameters are related to the evolution of galaxies \citep{conselice2003}.
The concentration index is correlated with the bulge-to-total ratio and stellar mass of the galaxies, while the clumpiness showed a strong correlation with 
H$\alpha$ equivalent width, which is indicative of the star-forming activity. 
The CAS system has recently been used to study the galaxy structure and its evolution at high redshifts with data obtained with the James Webb Space Telescope \citep{kartaltepe2023}. 
Although there are various studies using the CAS parameters in numerical simulations, 
they mainly focused on the merging process of the galaxies \citep[e.g.,][]{conselice2006, lotz2008, lotz2010a,lotz2010b}.

In this work, we proceed on the investigation of the disk morphological evolution starting from various Toomre's $Q$ parameter. Apart from the inspection of the conventional bar modes, we enlarge our scope to the CAS properties and how they can be relevant to observations. These parameters are widely employed to examine the structural properties of observed galaxies, and we will adopt them to analyze the numerical simulations in this study. The article is organized as follows. First, Sec. \ref{nume_acc} describes the numerical models and accuracy controls. Next in Sec. \ref{param}, we introduce the parameters used in this work, including those for probing the global non-axisymmetric features and the CAS indices. Then, Sec. \ref{nume} reports the numerical results and discussions are provided therein. Finally, Sec. \ref{conclu} concludes this study.

\section{Numerical simulations and accuracy controls}
\label{nume_acc}

The self-gravitating $N$-body simulations are handled by GADGET-2 \citep{springel+yoshida+white_2001,springel_2005}. The density profile of a disk of particles follows the exponential profile with vertical distribution that reads
\begin{equation}
\rho_{d}(r,z)=\frac{M_{d}}{4\pi R_{0}^{2}z_{0}}e^{-r/R_{0}}\text{sech}^{2}{\bigg(\frac{z}{z_{0}}\bigg)}
    \label{density_disk}
\end{equation}
where $M_{d}$ is the disk mass, $R_{0}$ is the disk scale radius, and $z_0$ is the disk scale height. We choose $M_{d}=10^{11} \ M_{\odot}$, $R_{0}=5 \ \text{kpc}$, and $z_{0}=0.2 \ \text{kpc}$. The disk is radially and vertically truncated at $5R_{0}$ and $5z_{0}$, respectively. We principally investigate the disks of $5\times 10^{6}$ particles, unless otherwise specified. To properly imitate the disk evolution in a dark matter halo, the disk is put in static spherical Hernquist potential given by
\begin{equation}
  \Phi_{h}(r)=-\frac{GM_{h}}{r+r_{h}}
  \label{pot_hern}
\end{equation}
where $M_{h}$ and $r_{h}$ are the halo mass and the halo scale radius, which are fixed to $2.5\times 10^{12} \ M_{\odot}$ and $75 \ \text{kpc}$, respectively.
The radial $Q$ profile corresponds to the ratio of the radial velocity dispersion $\sigma_{r}$ to the minimum required value for the local stability according to the Toomre's criterion \citep{toomre_1964}, i.e.,
\begin{equation}
  Q=\frac{\sigma_{r}\kappa}{3.36G\Sigma}
  \label{q_def}
\end{equation}
where $\kappa$ is the epicyclic frequency calculated from the composite disk-halo potential $\Phi_{tot}$. In the expression (\ref{q_def}), $\Sigma$ is the disk radial surface density which is proportional to $e^{-r/R_{0}}$. To construct a disk of particles in dynamical equilibrium with a static halo potential, we adopt the prescriptions of \citet{hernquist_1993} for the disk velocity structures. The squared radial velocity dispersion is proportional to the disk surface density, namely
\begin{equation}
  \sigma_{r}^{2}\propto e^{-r/R_{0}}. \label{sigmar_prop}
\end{equation}
The constant of proportionality in Eq. (\ref{sigmar_prop}) is adjusted so that the $Q$ value calculated by Eq. (\ref{q_def}) at $2R_{0}$, i.e., the reference radius, is equal to a chosen $Q$ value, and this $Q$ is the representative $Q$ for a case. In other words, a specific $Q$ can be obtained by properly adjusting the constant of proportionality in Eq. (\ref{sigmar_prop}). The tangential velocity dispersion $\sigma_{\theta}$ is obtained by the relation
\begin{equation}
  \sigma_{\theta}^{2}= \frac{\kappa^{2}}{4\Omega^{2}}\sigma_{r}^{2} \label{sigmatheta}
\end{equation}
where $\Omega$ is the angular frequency of circular orbit calculated from $\Phi_{tot}$. The vertical velocity dispersion $\sigma_{z}$ relates to $\Sigma$ as
\begin{equation}
  \sigma_{z}^{2}= \pi Gz_{0}\Sigma. \label{sigmaz}
\end{equation}
By these choices, we have $\sigma_{z}^{2}\propto\sigma_{r}^{2}$. The mean tangential velocity $\bar{v}_{\theta}$ as a function of radius is obtained by the axisymmetric Jeans equation as follows 
\begin{equation}
  \bar{v}_{\theta}^{2}=r\frac{d\Phi_{tot}}{dr}+\frac{r}{\Sigma}\frac{d(\sigma_{r}^{2}\Sigma)}{dr}+\sigma_{r}^{2}-\sigma_{\theta}^{2},
  \label{vthetamean}
\end{equation}
whereas the mean radial and vertical velocities (or $\bar{v}_{r}$ and $\bar{v}_{z}$, respectively) are initially zero. The three random velocity components are drawn from the cut-off Gaussian distribution with corresponding local velocity dispersion ellipsoid.

Calculations of spline-softened mutual forces for disk particles are facilitated by the tree code. We adjust the opening angle to $0.7$ and the softening length to $5 \ \text{pc}$ for all particles. The integration time step is controlled to be not greater than $0.2 \ \text{Myr}$. The accuracy is such that the deviations of both the total energy and the disk angular momentum at the end of simulation, which is specified to be $9.6 \ \text{Gyr}$, from the initial values are not greater than $0.1 \%$. We examine the cases where $Q=0.65, 0.8, 0.95, 1.1, 1.2$ and $1.5$.

\section{Parameters}
\label{param}

\subsection{Bar parameter}
\label{bar_param}
Non-axisymmetric features of a disk can be evaluated by the $m$-mode Fourier amplitudes as a function of radius $\tilde{A}_{m}(r)$ defined as
\begin{equation}
    \tilde{A}_{m}(r)=\frac{\sqrt{a^{2}_{m}+b^{2}_{m}}}{A_{0}}
    \label{tilde_a2}
\end{equation}
where $a_{m}$ and $b_{m}$ are the $m$-mode Fourier coefficients calculated from particles inside the annulus of radius $r$ and $A_{0}$ is the corresponding $m=0$ amplitudes. The $m$-mode strength $A_{m}$ is designated by the maximum $\tilde{A}_{m}$ within $r_{max}$, or  
\begin{equation}
    A_{m}\equiv \max_{r<r_{max}}[\tilde{A}_{m}]
\end{equation}
where we fix $r_{max}$ to $10 \ \text{kpc}$ for all calculations. We employ $A_{2}$ as the bar strength.

\subsection{CAS parameters}
\label{cas_param}

CAS system is a non-parametric measurement of light distribution of a galaxy. 
In the simulations, we obtain the surface density instead of the surface brightness. So we assume constant mass-to-light ratio across the entire simulated galaxy to translate the surface density into the surface brightness. 
It consists of 3 parameters: concentration ($C$), asymmetry ($A$), and clumpiness ($S$). 
We adopt the concentration index originally defined by \cite{bershady2000} as 
\begin{equation}
    C = 5 \log_{10} \left(\frac{r_{80}}{r_{20}}\right),
    \label{eq:con}
\end{equation}
where $r_{20}$ and $r_{80}$ are the radii governing 20\%
and 80\% of the growth curve within 1.5 times the Petrosian radius at $r(\eta=0.2)$, respectively. 

The asymmetry index shows the fraction of galaxy component that is not symmetric, which was originally defined as 
\begin{equation}
    A = \min\left(\frac{\sum  \lvert I_0-I_{180}\rvert}  {\sum \lvert{I_0}\rvert}\right) - \min \left(\frac{\sum\lvert B_0-B_{180}\rvert}{\sum\lvert I_0\rvert}\right),
    \label{eq:asym}
\end{equation}
where $I_0$ and $B_0$ are the original image and background area close to the galaxy, respectively \citep{conselice2000}. 
$I_{180}$ is the image after rotating $180^\circ$ from 
the center of the galaxy in the line of sight. 
The center is varied until we obtain the minimum value of the asymmetry index. To reduce the effect of noise, 
\cite{conselice2000} consider the background region near the object by randomly selecting its center, rotating the background by $180^\circ$, and calculating the rightmost term of Equation \ref{eq:asym}. 
However, we do not have noise in the simulation. 
So we simply use only the first term of Equation \ref{eq:asym}
to estimate an asymmetry index. That is 
\begin{equation}
    A = \min\left(\frac{\sum  \lvert I_0-I_{180}\rvert}  {\sum \lvert{I_0}\rvert}\right).
    \label{eq:asym2}
\end{equation}

Lastly, the clumpiness index is defined as the ratio of light in high-frequency structure to the total light of the galaxy \citep{conselice2003}. 
It can be written as 
\begin{equation}
    S = 10 \times \left[\left(\frac{\sum (I_{xy}-I^\sigma_{xy})}{\sum I_{xy}}\right) - \left(\frac{\sum (B_{xy}-B^\sigma_{xy})}{\sum I_{xy}}\right)\right],
    \label{eq:clumps}
\end{equation}
where $I_{xy}$ is is the original image, while $I^\sigma_{xy}$ 
is the smoothed image. Likewise, $B_{xy}$ and $B^\sigma_{xy}$ are, respectively, 
the original and blurred background with equal area to the size of the galaxy  \citep{conselice2014}. 
The original image is smoothed with a two-dimensional Gaussian kernel with the size of $\sigma=0.3\times r(\eta=0.2)$ \citep{conselice2003}.

\section{Numerical results}
\label{nume}

\subsection{Evolution of bi-symmetric modes}
\label{sim_overall}

\begin{figure}
\begin{center}
\includegraphics[width=8.0cm]{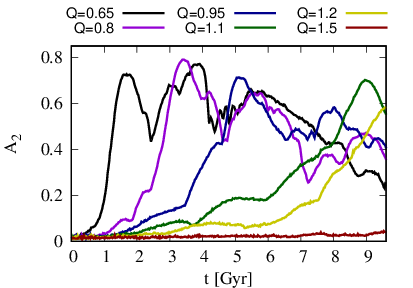}
\end{center}
\caption{Time evolution of $A_{2}$ for disks with different $Q$.}
\label{fig_a2_q}
\end{figure}

\begin{figure*}
\begin{center}
\includegraphics[width=16.0cm]{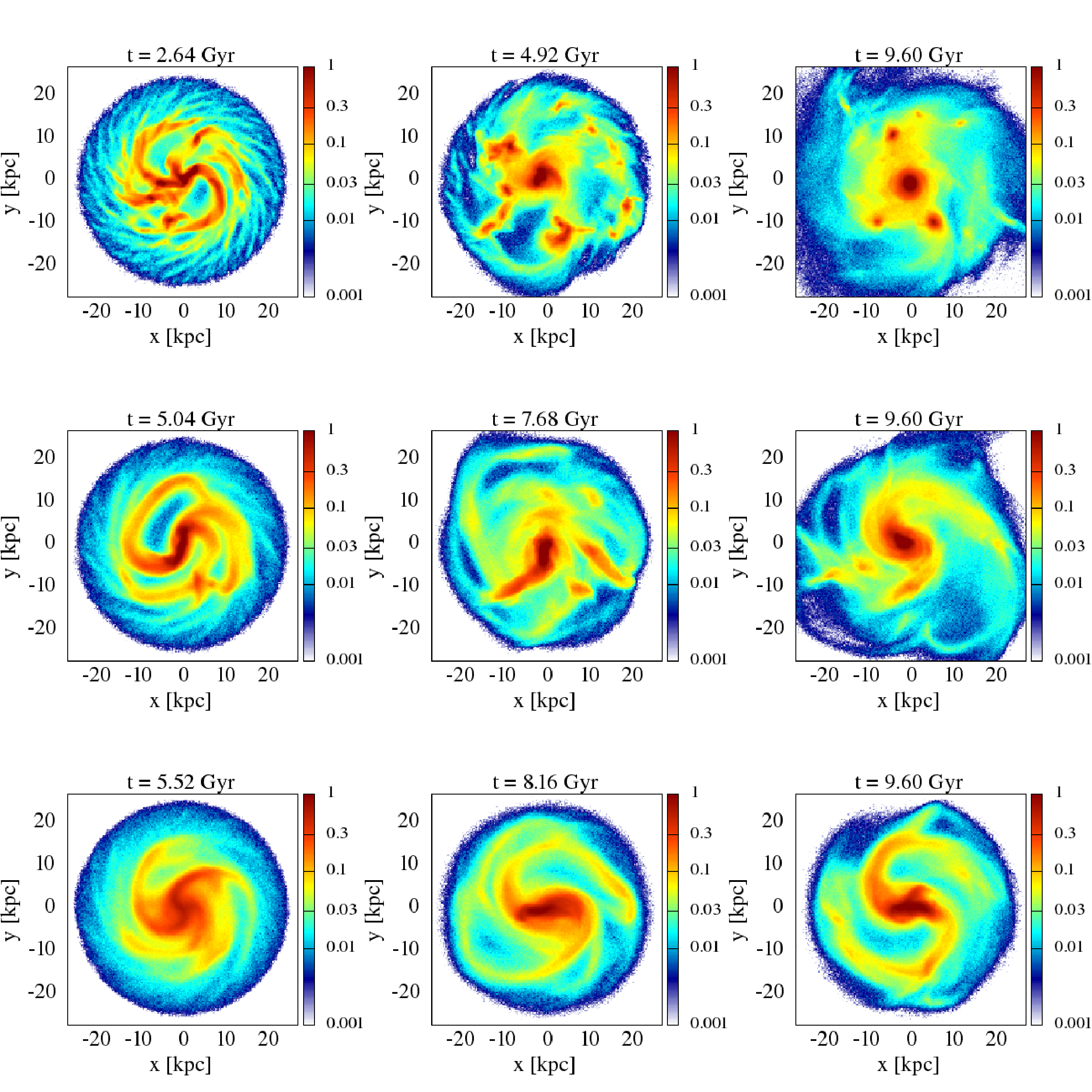}
\end{center}
\caption{Disk surface density map for $Q=0.65$ (top row), $0.95$ (middle row) and $1.1$ (bottom row) at indicated times.}
\label{fig_snap_q}
\end{figure*}

First of all, we inspect the time evolution of $A_{2}$ for disks with different $Q$ in Fig. \ref{fig_a2_q}. It turns out that the development of the bar modes takes place when $Q\leq 1.2$. Despite that $Q$ is less than $1$, the $m=2$ modes can still grow in the midst of the local instabilities. A disk with lower $Q$ tends to form a bar more rapidly, in accordance with past studies \citep{athanassoula+sellwood_1986,hozumi_2022}. On the other hand, the value of $Q=1.5$ is sufficient to suppress the bar instability. We further examine the disk configuration for different $Q$ in Fig. \ref{fig_snap_q} to verify the growth of the bar modes. In the case where $Q=1.1$, we observe the buildup of weak multi-armed modes at $5.52 \ \text{Gyr}$ which are overseen by the swing amplification \citep{julian+toomre_1966,toomre_1981}. That mechanism engenders the spiral arms from the synchronized epicyclic motion in combination with the disk shearing. After $8 \ \text{Gyr}$, the two-armed modes exert the dominance over the multi-arm modes as the configuration takes the form of the two-armed barred structure, and it remains in shape until the end.  
When $Q=0.95$, which is slightly below the local stability threshold, we observe a more rapid formation of the bar amidst the local instabilities as the bar is spotted at $5.04 \ \text{Gyr}$. Compared to the previous case, we remark clumpy sub-arms around the central bar. As time progresses, the shape of the bar and spiral arms deviates from bi-symmetry and they are surrounded by long-lasting clumps which remain until the end. 
The clumps emerge from the gravitationally unstable fluctuations, so they are distributed randomly in the disk environment. We speculate that the breaking of bi-symmetry is attributed to those local instabilities and the bar evolution in the midst of such instabilities can be described as follows. While the two-armed modes are growing, the unstable local over-densities progress to clumps in random places in parallel. Afterward, the bi-symmetric component is deformed by the interactions with clumps as observed in the last snapshot. 

The case where $Q=0.65$ is fully governed by local gravitational instabilities which lead to a different morphological evolution from the other cases. The mass concentrations develop quickly, supposedly in the time scale of the local free-fall time, before they are sheared by disk differential rotation. As a consequence, the multi-armed spiral pattern is formed rapidly within $\sim 2.6 \ \text{Gyr}$ which is significantly shorter than the time scale of the bar formation. The spiral arms appear more clumpy than those arising from the unstable two-armed modes or the swing amplification when $Q>1$. These arms are short-lived as they are found fragmented into clumps at $4.92 \ \text{Gyr}$. The clumps are still spotted around the concentrated disk center at the end of simulation. The fragile spiral structure is another indication of the dominance of the local instability over the spiral mode instability, which plays an important role in other cases. The plot of the density map informs that, referring to Fig. \ref{fig_a2_q}, the growth of $A_{2}$ does not signify the development of the bar-like structure. The developed $m=2$ mode amplitudes are caused by the anisotropically distributed clumps. We note that the number of the clumps in the end reduces from the number in the snapshot before which is possibly due to either the merger between them or the falling to the disk center. The decrease of the clump number is in accordance with the decline of $A_{2}$ with time near the end. 
In addition, we note the breaking of disk circular symmetry by clumps, which can be explained by the same arguments for the breaking of the bi-symmetry of the barred-spiral structure in the $Q=0.95$ case.

The bar formation in a disk with $Q\geq 1$ conforms with past mainstream studies but only a few studies visited the regime where $Q$ was below $1$ \citep{athanassoula_2003,debattista_et_al_2006,worrakitpoonpon_2023}. We have some important remarks for the latter regime that deserve attention. Our simulation of $Q=0.95$ disk demonstrates that, firstly, the disk is populated by long-lasting clumps that persist until the end. Furthermore, the overall symmetry is strongly broken compared with the $Q=1.1$ counterpart. These anomalies are attributed to the initial local instability. About the evolution pattern of the $Q=0.65$ disk, there was a conjecture for a disk of comparable $Q$ to be the progenitor of lenticular (or S0) galaxies which do not host the bar and the spiral arms \citep{saha+cortesi_2018}. That hypothesis is partly plausible for us as our simulations suggest that the S0 galaxy candidate can also emerge from $Q=1.5$, i.e., when the disk is stable against all types of perturbations. The differentiation between the practically similar configurations originating from the locally stable and unstable disks needs a suitable indicator. These related properties will be addressed in Sec. \ref{sim_cas}.

\subsection{Evolutions of CAS parameters}
\label{sim_cas}

\begin{figure*}
    \centering
    \includegraphics[width=0.9\textwidth]{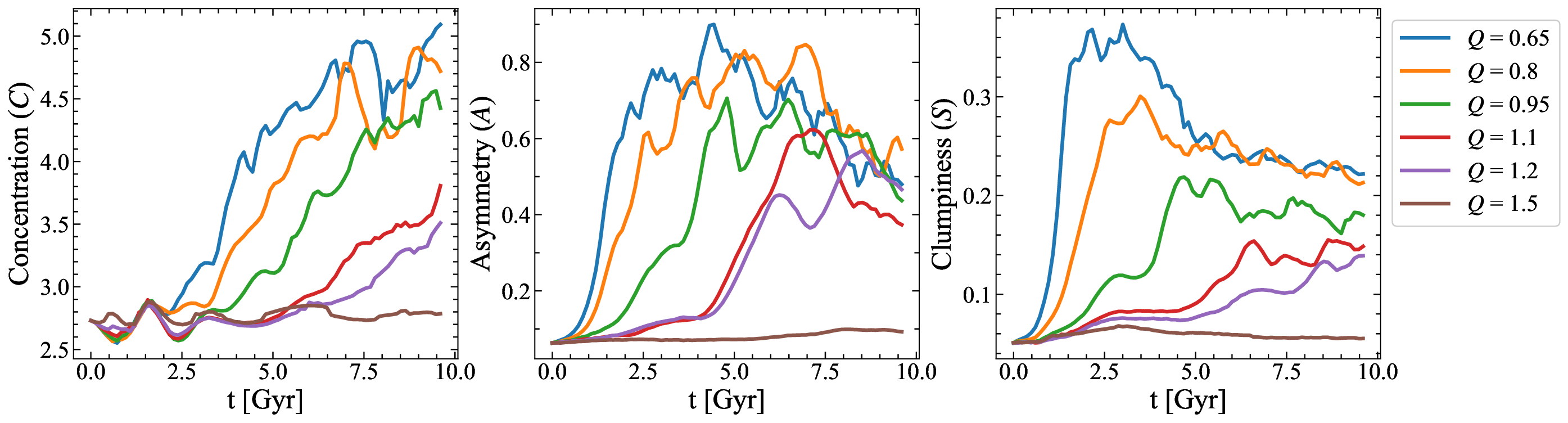}
    \caption{Time evolution of CAS parameters from the simulation with $N=5\times 10^6$. Different colors in all panels represent different values of $Q$ ranging from $Q=0.65$ to $Q=1.5$.}
    \label{fig:cas-time}
\end{figure*}

In continuity with the clumpy and asymmetric disks for $Q<1$ observed in Sec. \ref{sim_overall}, we continue with a more systematic evaluation of such properties. Shown in Fig. \ref{fig:cas-time} are the time evolutions of CAS parameters for different $Q$. From the plots, we are able to classify the evolution schemes into 3 sub-families: the $Q<1$ disks, the $Q>1$ disks that form a bar, and the disks that are bar-stable. 
Those with $Q<1$ develop highly concentrated, asymmetric, and clumpy features as remarked by large increases of the three parameters. With a closer look, the clumpiness $S$ increases most rapidly because of the local gravitational instabilities residing in the initial state. It remains at peak value for $\sim 2 \ \text{Gyr}$ before it decays sharply. Afterward, it keeps decaying slowly until the end but it is still higher than the other families. When $Q$ is closer to $1$, the $S$ index develops more slowly and reaches a lower peak. This is because the disk is closer to the local stability threshold. 
The decline of $S$ after the peak can be explained, in parallel with the evolution in Fig. \ref{fig_snap_q}, that the number of clumps reduces with time. We recall that the case where $Q=0.65$ undergoes the formation of clumpy multi-arm pattern in midway, in accordance with the time at which $S$ is at peak. In the end, a high level of clumpiness still retains, although the clumpy spiral arms dissolve. 
Considering the concentration $C$ and asymmetry $A$, they reach the peaks well after the $S$ peak. This indicates that the developments of $C$ and $A$ features are the secondary processes induced by the clumpiness that develops earlier. This order of increases is in accordance with earlier speculation that anisotropically distributed clumps are the source of asymmetry. Here, we validate that the enhanced concentration is also the consequence. Considering the evolution of $A$ index, the values at peak for $Q=0.65$ and $0.8$ are comparable and they are significantly higher than that for $Q=0.95$. After the peaks, the $A$ index for all three subcritical cases decays down. The rapid decay is explained that some clumps fall to the disk center, resulting in the enhancement of $C$ and the decline of $S$ accordingly. 
The $C$ parameter, on the other hand, does not show any sign of decay with time until the end of simulation, which indicates the continuous accumulation of the density at the center by merger with clumps. 
In summary, the CAS evolution of this disk family is dominated by local instabilities marked by rapid increase of $S$, followed by the increases of $C$ and $A$. It is worth reminded that the case where $Q$ is far below $1$ does not develop the bar-like structure, while those starting with $Q$ slightly below $1$ can develop a bar. These differences cannot be distinguished by using only the CAS parameters. A direct visual inspection on the disk configuration is required.

\begin{figure*}
    \centering
    \includegraphics[width=0.9\textwidth]{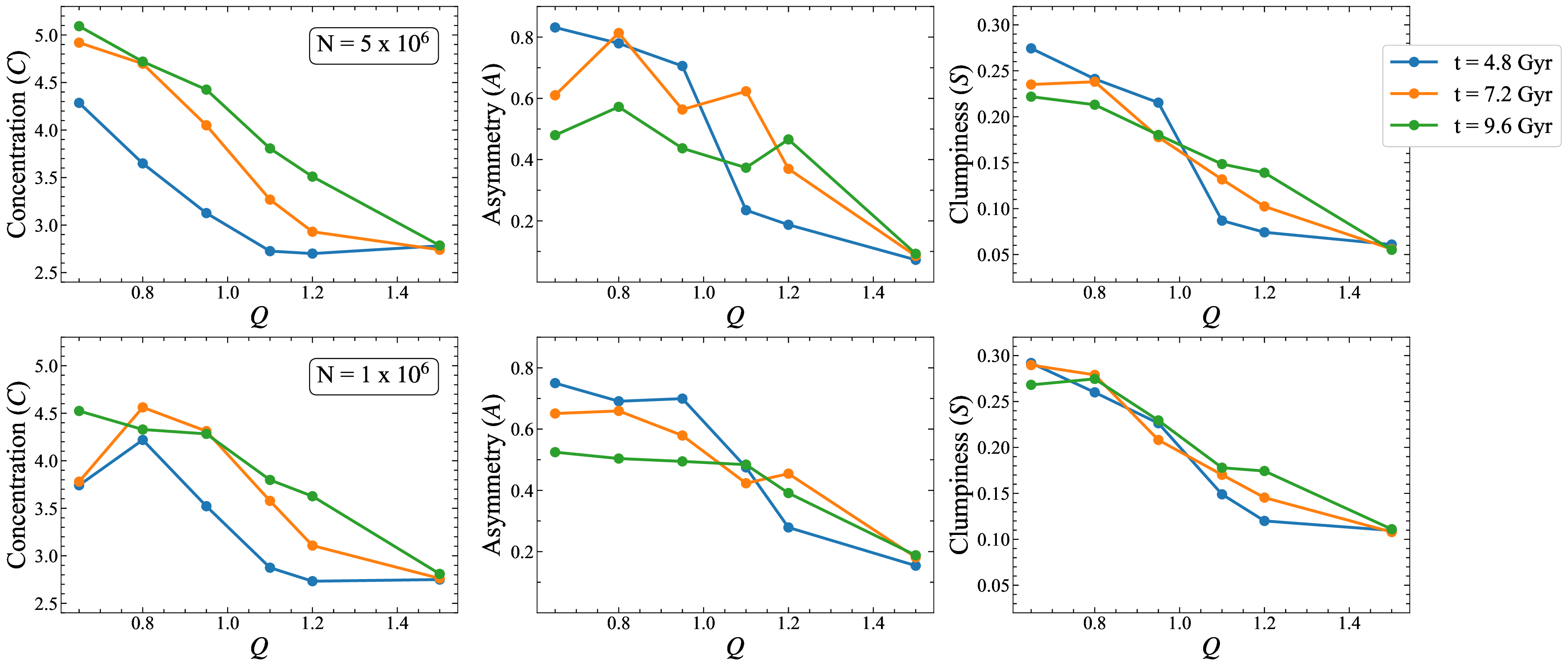}
    \caption{CAS parameters as a function of $Q$ for $N=5\times10^6$ (top panels) and $N=1\times10^6$ (bottom panels). Blue, orange, and green solid lines indicate the snapshots at $t=4.8$ Gyr, $t=7.2$ Gyr, and $t=9.6$ Gyr, respectively. Solid circles represent the values of $Q$ we adopted in each simulation.}
    \label{fig:cas-q}
\end{figure*}

Disks that are bar-unstable and start with $Q>1$ exhibit moderate increases of all CAS parameters and the order of increases differs from those for $Q<1$. It is the $A$ parameter that increases first due to the growth of two-armed modes that finally become the barred-spiral structure, and that established structure induces the growths of $C$ and $S$. In this disk family, the linearly unstable two-armed modes play a central role unlike the situation in the previous disk family. 
Although the local gravitational instabilities are suppressed, $C$ and $S$ can nevertheless be enhanced by the bar and the two-armed modes, in line with the concentrated bar and spiral arms observed at the end in Fig. \ref{fig_snap_q}, but they are not at the levels attained in a locally unstable disk. Close to the end of simulations, the $A$ index decreases for $Q=1.1$, which marks the onset of the decay of bar modes in coherent with the decrease of $A_{2}$ in Fig. \ref{fig_a2_q}. The case where $Q=1.2$ does not yet exhibit a decay because the bar is formed later. In the bar-stable family, all parameters remain close to initial values. This indicates that both local and bar-mode instabilities are effectively suppressed.

The variation of CAS parameters can otherwise be plotted as a function of $Q$ as shown in Fig. \ref{fig:cas-q} at different times for $N=5\times 10^{6}$ and $10^{6}$. The investigation of the case with a lower $N$ has a purpose to examine how the results change by reducing $N$ since the clumps originate from the gravitationally unstable finite-$N$ fluctuations, whose amplitudes scale with $1/\sqrt{N}$. 
In both cases, we observe that $C$ and $S$ tend to decrease as $Q$ increases at $9.6 \ \text{Gyr}$, as expected from a lesser degree of local instabilities. With a closer look, varying $N$ causes some notable differences in evolutionary pattern and magnitude. The trivial consequence is that the $S$ parameter for $N=10^{6}$ is averagely higher because of more pronounced local instabilities. However, the decrease with $Q$ retains for both $N$. For the $C$ parameter, it monotonically decreases with $Q$ in all time snapshots for $N=5\times 10^{6}$. When $N$ steps down to $10^{6}$, the value for $Q=0.65$ is significantly lower than the value for $Q=0.8$ during $4.8-7.2\ \text{Gyr}$, but the monotonic decrease with $Q$ is eventually achieved at $9.6 \ \text{Gyr}$. This can be explained that the developed clumpy feature lasts longer if $N$ is lower before it falls to the disk center. This speculation is supported by a slower decay of $S$ index from $4.8-9.6 \ \text{Gyr}$ than the decay rate for $N=5\times 10^{6}$.
For the $A$ parameter, the decrease with $Q$ is more evident in the middle of evolution because of the existing clumps and the forming barred-spiral structure whereas there is no clear tendency of the variation with $Q$ at the end. It turns out that using the $A$ index as an indicator of the initial $Q$ is at best when non-axisymmetric features are at peak. The variation of $A$ with time and with $Q$ for $N=10^{6}$ is qualitatively similar to the $N=5\times 10^{6}$ counterpart: the decrease with $Q$ is obvious at $4.8 \ \text{Gyr}$ but the curve becomes flattened in the range of $Q\leq 1.1$ in the end. In addition, the decay of $S$ results in the decrease of $A$ as some clumps are falling to the symmetric disk center and become part of it.

The inspection of CAS parameters provides a new insight in the way that the origin and evolution history of a disk of any morphology and structural detail can be specified. In the bar-forming regime with $Q$ ranging from $0.8$ to $1.2$, disks starting with $Q$ below and above $1$ exhibit distinct evolutionary tracks of CAS indices and they lead to significantly different values. More specifically, barred galaxies formed in $Q<1$ disks can be identified by $C>4.0$ and $S>0.15$, referring to the case of $N=5\times 10^{6}$, and vice versa for bar-forming disks starting with $Q>1$. Fig. \ref{fig:cas-time} suggests that the proposed $C$ border value is applicable in the time frame of galaxy age from $7-9.6 \ \text{Gyr}$. The differentiation of the disk origin by that proposed $S$ border value is applicable to a wider time frame of galaxy age: the value of $S=0.15$ can distinguish the initial kinematical conditions of disk galaxies of age from $5-9.6 \ \text{Gyr}$. On the other hand, the $A$ value can be useful when the asymmetric feature is at peak. The clumpy barred-spiral disk can produce $A$ that is significantly greater than $0.6$ and it saturates on that level for more than $4 \ \text{Gyr}$. This is significantly greater than the asymmetry caused solely by the barred-spiral pattern, which can attain the value around $0.6$ at most for a short period of $\sim 0.5 \ \text{Gyr}$ before it starts to decay. The $A$ parameter can therefore be utilized in the following way. The observed value of $A$ greater than $0.6$ can be an evidence of the locally unstable initial disk. Otherwise, the establishment of the barred-spiral galaxies without local instabilities cannot yield an $A$ value greater than $0.6$ in most of their lifetime.

Considering unbarred cases, this configuration can be achieved when $Q$ is either far above or far below $1$. Both cases yield the lenticular-like morphology at the end but the measured CAS parameters are at remarkably different levels. The values of $C>4.5$, $A>0.4$, and $S>0.2$ can infer a cold progenitor of a lenticular disk, which undergoes violent evolution forming clumpy short-lived spiral arms. Orbiting clumps that remain are another piece of evidence of cold origin. On the contrary, those with modest CAS magnitudes undergo smoother evolution with no perturbative structures formed because all local and global non-axisymmetric modes are suppressed. The $C$, $A$ and $S$ values can be as low as $2.7$, $0.1$ and $0.05$, respectively.

In observations, the concentration index increases with the increasing bulge-to-total ratio for the local galaxies \citep{conselice2003}. The early-typed disk and elliptical galaxies, which the bulge is already well established, showed the concentration index of $C>4.0$. Although this criterion is consistent with the simulated barred galaxies with $Q<1$ in our simulations, the concentration index alone cannot be efficiently used to indicate the bar structure. 
The asymmetry index is famously used to identify the merging system in both observations and simulations. \cite{lotz2010a} used the simulations to show that the asymmetry index could reach $A>0.4$ at the stage of final merger. 
It has been later used as the criterion for merging system in the observations \citep{kartaltepe2023}. 
Considering our numerical results, such high level of the $A$ index can be achieved and maintained for long period in cases with $Q<1$, whereas that level can be attained momentarily during the peak of the bar modes when $Q>1$. Our important finding is that this high level of asymmetry can be attained spontaneously in locally unstable disks. 
In case of an isolated galaxy, the clumpy object can also show high asymmetry index, 
as it is strongly correlated with the clumpiness \citep{conselice2003}. 
This correlation is also seen in our simulations (Fig. \ref{fig:cas-time}). 
Early-type spiral and elliptical galaxies showed very low asymmetry ($A<0.1$) and clumpiness ($S<0.1$), while the late-type spiral or starburst with high star formation rate show the larger values. 
The enhanced star formation in galaxies that exhibit a higher clumpiness index is actually understandable as the formation of the clumps also promote the local star formation.
This observed correlation emphasizes the possibility of the locally unstable disk to be a progenitor of clumpy spiral galaxy.

In summary, we find a utility of CAS parameters apart from their  correlations with some present physical properties of galaxies such as the bulge-to-total ratio and the star formation rate. The CAS magnitudes of an established disk galaxy of any morphology can determine its origin and evolution history reasonably well. 
Although our numerical model consists of purely self-gravitating particles, whereas a real galaxy consists of many components, the conclusion of the relation between the CAS values and the evolutionary history remains applicable. This is because the key factor leading to the difference, namely the local instability, is overseen by the gravity and the gravity is the governing force of all components. In addition, it was documented that observed disk galaxies with estimated $Q$ less than $1$, in which local instabilities played a central role, tended to be gas-poor \citep{renaud_et_al_2021,aditya_2023}. This ascertains the dominance of the gravitational dynamics, which is our focus here, over the gas dynamics for this $Q<1$ regime. 
Another limitation is that our study considers the evolution of a disk in isolation. A galaxy with merging history might yield the elevated $A$ without local instabilities that might not conform with our conclusion.

\subsection{Radial heating with and without clumps}
\label{sim_kine}

\begin{figure}
\begin{center}
\includegraphics[width=8.0cm]{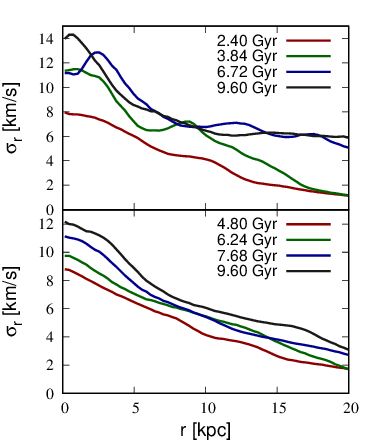}
\end{center}
\caption{Radial velocity dispersion profiles at different times for $Q=0.8$ (top panel) and $1.1$ (bottom panel).}
\label{fig_dispr}
\end{figure}

It has been justified that the evolving non-axisymmetric forces from the bar and the spiral arms could radially heat the disk environment, i.e., the radial heating \citep{jenkins+binney_1990,minchev+quillen_2006}. Shown in Fig. \ref{fig_dispr} is the radial velocity dispersion profile for $Q=0.8$ and $1.1$ at different times covering the process of the bar formation and its subsequent evolution. In both cases, we capture the radial heating as the profile raises with time, but the detailed procedures are different. For $Q=1.1$, the entire profile heats up smoothly by global bi-symmetric forces. 
On the contrary, bumps in the profile of the $Q=0.8$ case during the heating are spotted, most notably at $3.84 \ \text{Gyr}$. The bumps are lowered but still observable at $6.72 \ \text{Gyr}$ before the profile is smoothed out at $9.60 \ \text{Gyr}$. We note that the final profile is flatter than the $Q=1.1$ profile at the same time. These differences are attributed to the additional local heating by clumps which is absent when $Q>1$, in which case the disk is heated by global bi-symmetric forces without local effect. The bumps in the profile during the radial heating and the flatness of the profile at the end can be indicative to the kinematically cold disk origin, in addition to the elevated CAS values that we addressed in the previous section.

The concept of the radial heating by clumps shares similarities with the framework of the heating by the giant molecular cloud (GMC) proposed by, for instance, \citet{hanninen+flynn_2002,aumer_et_al_2016,fujimoto_et_al_2023}. The major difference is, however, that those studies constructed the GMC in the context of the locally stable disk, i.e., $Q\geq 1$, while we inspect the locally unstable disk that spontaneously forms clumps. However, we do not rule out the relevance to that hypothesis as in reality, the disk kinematical map is not smooth and perfectly axisymmetric as imposed in the initial conditions for simulations. The GMC can potentially be formed in a location where local $Q$ is below $1$, although the disk has $Q\geq 1$ on average.

\subsection{Lopsided bar and spiral arms in subcritical $Q$ disks}
\label{sim_lop}

\begin{figure}
\begin{center}
\includegraphics[width=8.0cm]{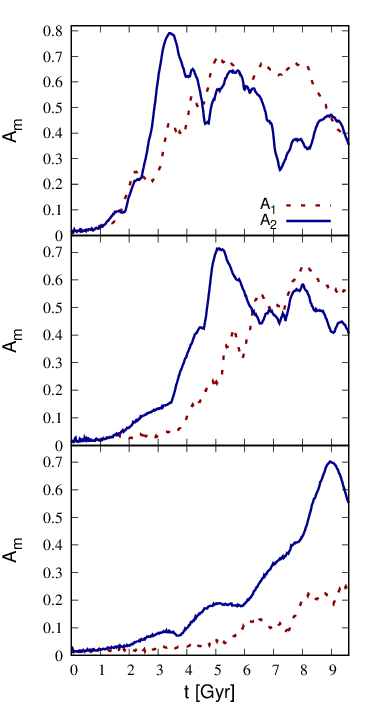}
\end{center}
\caption{Time evolutions of $A_{1}$ (dotted line) and $A_{2}$ (solid line) for $Q=0.8$ (top panel), $0.95$ (middle panel) and $1.1$ (bottom panel).}
\label{fig_a1a2}
\end{figure}

\begin{figure}
\begin{center}
\includegraphics[width=8.0cm]{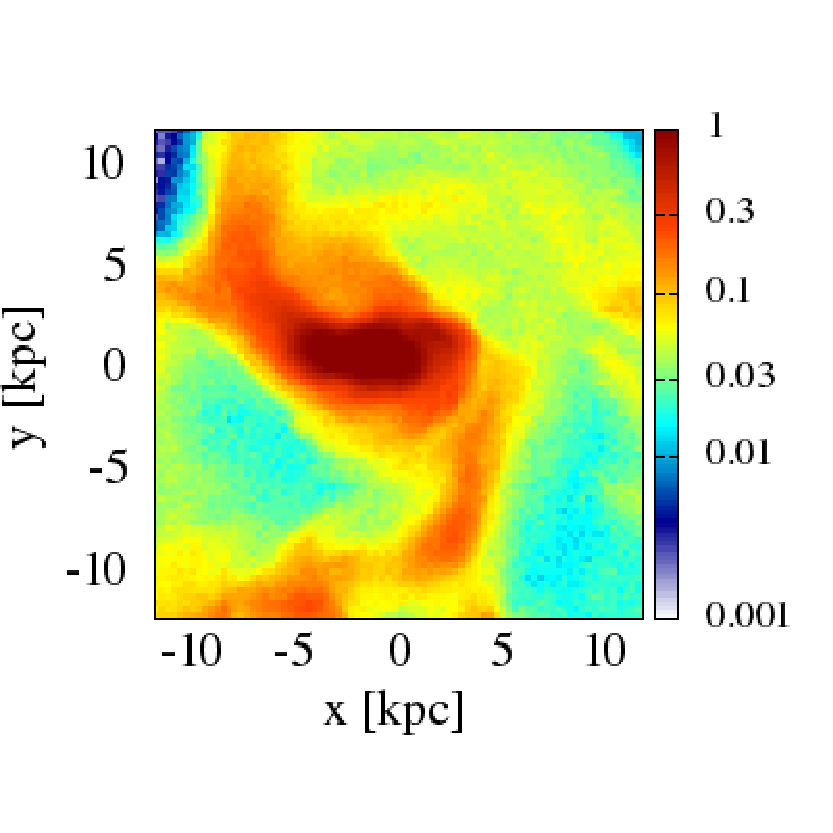}
\end{center}
\caption{Surface density map of disk center for $Q=0.95$  at $6.96 \ \text{Gyr}$.}
\label{fig_snap_lopsided}
\end{figure}

In continuity with the notice of an apparently lopsided bar for the case where $Q=0.95$ in Sec. \ref{sim_overall}, here we carry out a more systematic inspection of that appearance. It is a usual practice to probe the lopsidedness by $A_{1}$ and we will inspect its evolution in time along with $A_{2}$. Shown in Fig. \ref{fig_a1a2} are time evolutions of $A_{1}$ and $A_{2}$ for various $Q$. In all cases, the $m=2$ modes are growing faster and they lead to the barred structure when those modes are at peak.
Afterward, the $m=1$ modes are growing whereas the $m=2$ counterparts are decaying for cases with $Q=0.8$ and $0.95$ and the $m=1$ modes become dominant at the end. The development of $m=1$ modes relative to the growth of $A_{2}$ for $Q=0.95$ is slower than in the $Q=0.8$ case because of less pronounced local effect.
The emerging lopsidedness as speculated by the dominance of $A_{1}$ is validated by Fig. \ref{fig_snap_lopsided} which illustrates the lopsided barred-spiral disk around the time of $A_{1}$ peak for $Q=0.95$.
The evolution schemes of $A_{1}$ and $A_{2}$ affirm our hypothesis in Sec. \ref{bar_param} that anisotropically distributed clumps cause the breaking of bi-symmetry. On the other hand, the $m=1$ modes are not much amplified after the growth of the $m=2$ modes when $Q=1.1$, and the latter modes remain the dominant modes until the end. This is because of the lack of the local instabilities.
 
That the lopsided bar can potentially be established in subcritical $Q$ regime is an alternative explanation because past studies often related the origin of lopsided feature to the imposed macroscopic asymmetry of the initial state. For instance, numerical results of \citet{levine+sparke_1998,lokas_2021lop} demonstrated that a disk that was initially off-centered from the halo origin could produce a lopsided bar. Other studies attributed the formation of lopsided bar to tidal interactions \citep{yozin+bekki_2014,lokas_2022lop,varela-lavin_et_al_2023}. There were hypotheses that other astrophysical processes such as the gas accretion \citep{bournaud_et_al_2005,dupuy_et_al_2019} and the stellar feedback \citep{manuwal_et_al_2022} could generate the lopsidedness. 
Our important finding is that the established bi-symmetry can spontaneously be broken and the disk becomes lopsided, although the initial disk is constructed from a circularly symmetric profile, if the gravitationally unstable fluctuations are taken into account. In our framework, the initial asymmetry and the external forces are not required.

\section{Conclusion}
\label{conclu}

We examine the evolution of disk galaxies in simulations for various initial $Q$ values covering those that are below the Toomre's stability threshold and those above it. The scope is not only limited to the evolution of bi-symmetric features, but we also investigate the development of concentration, asymmetry and clumpiness (CAS) properties in that range of $Q$. The CAS parameters are widely adopted to discriminate galaxies in observations and we will apply those parameters to our numerical results. 

We find a utility of CAS indices when they are considered along with other conventional plots such as the bar parameter or the density map as they are able to identify the kinematical condition of the initial disk and the evolution history. For instance, the measured CAS values of cases that form a bar, which cover the initial $Q\in[0.8,1.2]$, are able to distinguish them into two sub-groups. A disk starting with $Q<1$ develops a bar and clumps in parallel, yielding high CAS indices at the end, while a $Q>1$ counterpart develops only a bar with two-armed spiral pattern. In the latter group, the CAS values are significantly lower. For unbarred cases which are the candidates of the lenticular galaxy, this morphology can be obtained from an initial disk with $Q$ either far below or far above $1$. The discrimination of unbarred disks achieved from those initial states at the two extremities can also be done using the CAS indices. The former case yields considerably more elevated CAS values than those in the latter case. High CAS values in far subcritical $Q$ disks also reflect the clumpy and asymmetric appearance which is not complicated to be remarked visually.

In addition, the presence of long-lived clumps surrounding the barred-spiral structure can lead to remarkable outcomes. The first consequence is the distinct features of the radial velocity dispersion profile caused by the local radial heating by clumps. It gives rise to the bumps in the profile while the heating is progressing and the flatter profile at the end than that without local heating. Secondly, the interactions between clumps and bi-symmetric components are able to deform the bar to lopsided configuration. This is an alternative explanation of the formation of the lopsided bar apart from other scenarios which require an initial seed of asymmetry or external forces. 

\vspace{0.5cm}

This research has received funding support from the NSRF via the Program Management Unit for Human Resources \& Institutional Development, Research and Innovation (grant number B16F640076).  TC is supported by Naresuan University (NU), and the National Science, Research and Innovation Fund (NSRF) Grant number R2566B091.  TW has funding support from Thailand Science Research and Innovation (TSRI) via Suranaree University of Technology [grant number 179349]. SY is supported by Mahidol University (Fundamental Fund: fiscal year 2023 by National Science Research and Innovation Fund (NSRF)) and by office of the Ministry of Higher Education, Science, Research and Innovation through research grant for new scholars (RGNS63-175). Numerical simulations are facilitated by HPC resources of Chalawan cluster of the National Astronomical Research Institute of Thailand.


\end{document}